\documentclass[fleqn, usenatbib]{mnras}

\usepackage{graphicx}
\usepackage{textcomp}
\usepackage{amssymb}\usepackage{hyperref}
\usepackage{amsmath,alltt}           
                                                                
\usepackage{blindtext}
\usepackage{multirow}
\usepackage{rotating}
\usepackage{lscape}
\usepackage{times}
\usepackage{hyperref}
\usepackage{pdflscape}
\usepackage{supertabular}
\usepackage[bottom]{footmisc}

\DeclareRobustCommand{\VAN}[3]{#2}
\let\VANthebibliography\thebibliography
\def\thebibliography{\DeclareRobustCommand{\VAN}[3]{##3}\VANthebibliography}

\def\apgt{\ {\raise-.5ex\hbox{$\buildrel>\over\sim$}}\ }
\def\aplt{\ {\raise-.5ex\hbox{$\buildrel<\over\sim$}}\ }
\title[Mergers of Equal-Mass Binaries in Stellar Triples]{Mergers of Equal-Mass Binaries with Compact Object Companions from Mass Transfer in Triple Star Systems}
\author[Leigh N. W. C., Toonen S., Portegies Zwart S. F., Perna  R.]{Nathan W. C. Leigh$^{1,2}$, 
Silvia Toonen$^{3}$, Simon F. Portegies Zwart$^{4}$, Rosalba Perna$^{5,6}$
\thanks{E-mail: nleigh@amnh.org (NL)}\\
$^{1}$Departamento de Astronom\'ia, Facultad de Ciencias F\'isicas y Matem\'aticas, Universidad de Concepci\'on, Concepci\'on, Chile \\
$^{2}$Department of Astrophysics, American Museum of Natural History, New York, NY 10024, USA\\
$^{3}$Institute for Gravitational Wave Astronomy, School of Physics and Astronomy, University of Birmingham, Birmingham, B15 2TT, UK\\
$^{4}$Leiden Observatory, Leiden University, PO Box 9513, 2300 RA, Leiden, the Netherlands\\
$^{5}$Department of Physics and Astronomy, Stony Brook University, Stony Brook, NY, 11794, USA\\
$^{6}$Center for Computational Astrophysics, Flatiron Institute, 162 5th Avenue, New York, NY 10010, USA}

\begin{document}

\pagerange{\pageref{firstpage}--\pageref{lastpage}} \pubyear{2011}

\maketitle

\label{firstpage}

\begin{abstract}

In this paper, we consider triple systems composed of main-sequence (MS) stars, and their internal evolution due to stellar and binary evolution.  Our focus is on triples that produce white dwarfs (WDs), where Roche lobe overflow of an evolving tertiary triggers accretion onto the inner binary via a circumbinary disk (CBD) driving it toward a mass ratio of unity.  We present a combination of analytic- and population synthesis-based calculations performed using the \texttt{SeBa} code to constrain the expected frequency of such systems, given a realistic initial population of MS triples, and provide the predicted distributions of orbital periods.  We identify the parameter space for triples that can accommodate a CBD, to inform future numerical simulations of suitable initial conditions.  We find that $\gtrsim$ 10\% of all MS triples should be able to accommodate a CBD around the inner binary, and compute lower limits for the production rates.  This scenario broadly predicts mergers of near equal-mass binaries, producing blue stragglers (BSs), Type Ia supernovae, gamma ray bursts and gravitational wave-induced mergers, along with the presence of an outer WD tertiary companion.  We compare our predicted distributions to a sample of field BS binaries, and argue that our proposed mechanism explains the observed range of orbital periods.  Finally, the mechanism considered here could produce hypervelocity MS stars, WDs and even millisecond pulsars with masses close to the Chandrasekhar mass limit, and be used to constrain the maximum remnant masses at the time of any supernova explosion.

\end{abstract}

\begin{keywords}
stars: blue stragglers -- binaries: close -- accretion, accretion disks -- hydrodynamics -- supernovae: general -- stars: white dwarfs
\end{keywords}

\section{Introduction} \label{intro}

Compact objects (COs) are thought to be responsible for a large fraction of observational high-energy astrophysics phenomena. Most of the UV, X-ray and gamma ray photons come from some form of accretion onto COs, whether it be from a companion star over-filling its Roche lobe in a binary star system, a gas disk surrounding and accreting onto a black hole, or the tidal disruptions of stars by super-massive black holes, and so on.  The mergers of, and accretion onto, compact objects specifically are directly relevant to a number of sub-disciplines within astrophysics, such as the origins of Type Ia supernovae (SNeIa), gamma ray bursts (GRBs), low-mass X-ray binaries (LMXBs), millisecond pulsars (MSPs), cataclysmic variables, etc.  

The origins of Type Ia SN remain unknown, although there is a general consensus that SNeIa are the products of runaway thermonuclear explosions of degenerate carbon-oxygen white dwarfs (WDs) \citep[e.g.][]{wang12,maoz14,livio18,ruiter20}.  There are two classical channels for inducing such a runaway explosion.  In the ``single degenerate'' scenario \citep[e.g.][]{whelan73,nomoto84}, a carbon-oxygen WD accretes mass from a companion star until it exceeds the Chandrasekhar mass limit. In the ``double degenerate'' scenario \citep[e.g.][]{iben84,webbink84}, two WDs comprise a compact post-common envelope (CE) binary system that is driven to merger via the emission of gravitational waves (GWs).

Given that neither of the above scenarios have as of yet been able to reproduce the inferred event rates, other mechanisms for SNeIa have recently been proposed in the literature, and are gaining attention.  The WD-WD collision scenario is particularly interesting, in which two WDs in a binary gain a sufficiently high eccentricity that they collide nearly head-on.  An explosion is then triggered by the very high temperatures and shocks generated during the collision \citep[e.g.][]{benz89,raskin09,rosswog09}.  If a tertiary star orbits an inner compact WD-WD binary, then the required high eccentricities for the collision scenario can be achieved due to Lidov-Kozai oscillations \citep{lidov62,kozai62,naoz16}.  Contrary to early claims \citep{katz12}, although such triple-induced mergers likely contribute to the total observed SNeIa rate, \citet{toonen18} recently showed that it is likely a small fraction of the total since high mutual inclinations are needed between the inner and outer triple orbital planes to induce WD-WD collisions, and the stellar progenitors of these systems would likely have merged before WD formation.         

Modifications of the aforementioned triple scenario could help to increase this mechanism's contribution to the total overall rates.  Examples include quadruple configurations hosting additional orbital planes to interact via Lidov-Kozai oscillations in more complex ways \citep{hamers18,fang18}, field triples that become dynamically unstable and produce WD-WD collisions \citep{perets12}, gravitational perturbations from stars passing by the triples which can serve to increase the mutual inclination between the inner and outer orbits and/or the outer orbital eccentricity \citep{antognini16}, WD natal kicks due to asymmetric mass loss with a magnitude of typically 0.75 km s$^{-1}$ \citep{fellhauer03,hamers19}, and so on.  

Recently, \citet{dong15} reported the discovery of doubly peaked line profiles in 3 out of $\sim$ 20 SNeIa using high-quality nebular-phase spectra.  The two peaks consistently seen in the Co/Fe emission features are respectively blueshifted and redshifted relative to the host galaxy, with a separation of $\sim$ 5000 km s$^{-1}$.  Only a small fraction of SNe with intrinsically bimodal velocity distributions will appear as doubly peaked spectra, due to their random orientations relative to our line of sight.  Hence, the authors conclude that SNe with intrinsic bimodality are probably common.  They further argue that such WD-WD collisions are naturally produced in triples, and demonstrate the detonation of both WDs using 3D hydrodynamical simulations of 0.64+0.64 M$_{\odot}$ WD-WD collisions.  However, we note that, from stellar and binary evolution alone, it is extremely unlikely that such a high mass ratio near unity would be produced.

Despite the aforementioned promising suggestions to relieve current tensions between observational data and theory, a great deal about the actual physics of mass transfer remains largely unknown.  Hence, for example, instead of a direct collision causing an SNeIa, the more massive WD binary companion could accrete steadily from its WD companion, allowing it to accrete sufficient mass to detonate as a Type Ia SN.  In this scenario, the surviving (lower-mass) WD could potentially be ejected as a hypervelocity object\footnote{Classically, the term "hypervelocity" implies a velocity exceeding the local Galactic escape speed.  We use this term more loosely here, referring instead to objects with velocities $\gtrsim$ 500 km s$^{-1}$ in the rest frame of the Galaxy.}, leaving the system with an escape velocity roughly equal to its orbital velocity at the time of its companion's explosion.  In fact, three WD candidates were recently identified in the GAIA DR2 catalog and suggested to have hypervelocities \citep{shen18}.  

The contribution to Type Ia SN from WD-hosting triple star systems remains unknown, in spite of a myriad of additional physics that could be mediated by the presence of a tertiary companion \citep[e.g.][]{hamers13,toonen18}.  The same can be said of other types of stellar exotica, most notably blue stragglers \citep{perets09,leigh11} and cataclysmic variables \citep[e.g.][]{shara17,shara18}.  In the case of blue stragglers, for example, triples have been proposed to produce long-period BS-MS binary systems, either via Kozai-Lidov oscillation-induced mergers of the inner binaries of MS triples \citep{perets09} or collisions of MS stars via dynamical interactions involving triples that perturb the inner binary to merge \citep{leigh11}.  As we will show in this paper, a number of interesting production channels for these types of exotic systems could be facilitated via stellar evolution within triple systems.  

Arguably the currently available empirical data are most consistent with binary evolution being the dominant BS formation mechanism operating in not only sparse environments, such as the Galactic field and open clusters, but also much denser globular clusters \citep{portegieszwart97,leigh07,knigge09,leigh11,leigh13,mathieu09,gosnell14}.  This is instead of direct stellar collisions, which can occur commonly during direct, chaotic interactions involving single and binary stars \citep{portegieszwart97,leonard89,leigh11}.  In the mass transfer (MT) scenario, one component of a MS-MS binary evolves to overfill its Roche lobe, transferring mass to its MS companion.  In the end, the system becomes a rejuvenated MS star or BS in a most-likely long-period binary star system with an outer WD companion.   \citet{gosnell14} tested this hypothesis in the old open cluster NGC 188, using data taken from the \texttt{Hubble Space Telescope}.  The authors searched for hot, young WD companions to the $\sim$ 20 BSs observed in NGC 188.  They identified 2-3 hot, young WDs and argued that this detection frequency is consistent with \textit{all} BSs in NGC 188 having formed from binary mass transfer.  This is because WDs should cool and dim rapidly post-formation, quickly becoming too dim to be observable over the bright blue BS companion, which is thought to have a much longer MS lifetime than the time required for a WD to cool to the point of being invisible.  Although triples are known to exist in non-negligible numbers in all BS-hosting stellar environments \citep{leigh11,leigh13b}, and even some triples are known to themselves host BSs \citep{vandenberg01,sandquist03}, little work has been done to date to try to understand how mass transfer within triples could contribute to BS formation. 

Mass transfer is also thought to be crucial to the production of cataclysmic variables, along with (possibly recurring) novae and dwarf novae eruptions.  Here, mass is accreted from a MS companion in a compact binary star system onto a WD companion via an accretion disk.  WDs are notoriously unstable to mass transfer, and a thermonuclear runaway can be initiated on the WD surface if the MT rate is not very finely tuned, illustrated via simulations to lie around 10$^{-7}$ M$_{\odot}$ yr$^{-1}$ \citep[e.g.][]{2007ApJ...663.1269N,2017gacv.workE..56K,2013ApJ...777..136W,bours13,shara17,shara18,2019MNRAS.490.1678C,2020NatAs.tmp...61H}.  This prevents most of any accreted mass from being retained by the WD, such that it is unable to grow appreciably in mass.  Little work has been done to quantify the possible contribution of mass transfer in WD-hosting triples to the formation of cataclysmic variables and other related phenomena.

In this paper, we generalize the mechanism presented in \citet{portegieszwart19} to consider initial triple star systems composed of all main-sequence stars, and their subsequent evolutionary pathways due to stellar and binary evolution.  In particular, we consider an alternative modification of the triple scenario considered in \citet{portegieszwart19}, namely accretion onto an inner tight MS-MS, MS-WD or WD-WD binary from a circumbinary disk (CBD), being fed by a Roche lobe over-flowing evolved outer tertiary companion.  We naively expect the MS-MS, MS-WD or WD-WD binaries to accrete toward a mass ratio close to unity via the CBD for at least some subsets of the total available phase space.  Although detailed hydrodynamics simulations have not been able to identify any clear trends \citep[e.g.][]{artymowicz83,young15,miranda15,munoz19,mosta19}, 
this general behaviour can be expected when the secondary's Roche lobe is further from the system's centre of mass, and becomes more likely to accrete high angular momentum material from the CBD,  driving the mass ratio back toward unity, as found in \citet{young15}, \citet{portegieszwart19} and \citet{duffell19}.  Our ultimate goal in this paper is to identify the subset of the total parameter space corresponding to triples that can accommodate a CBD, in order to inform future studies of suitable initial conditions to adopt for hydrodynamics simulations.
We further consider a novel mechanism for the production of compact object (CO) binaries with equal mass ratios.  We discuss the relevance and implications of this mechanism for producing Type Ia SN, in addition to blue stragglers, cataclysmic variables and other types of stellar exotica and CO mergers, including those involving neutron stars and black holes.  

%

\section{Methods} \label{methods}

In this section, we discuss and quantify the formation of triples composed of an inner binary containing one or more white dwarfs, with a tight outer tertiary orbit.  The final outer orbit must be sufficiently tight to ensure mass transfer at some point before the tertiary also evolves into a compact object, with significant binary evolution in the inner binary occurring in the interim, which comes along with additional complications we quantify in this section.  In order to quantify the expected frequency of such triples, we use a combination of analytic methods and population synthesis-based calculations for binary evolution using the \texttt{SeBa} code \citep{portegieszwart96,toonen12}.

\subsection{The Progenitor Stellar Triple and its Evolution Toward Hosting COs} \label{prog}

In this section, we consider the progenitor stellar triple composed of all MS stars, and the implications of any common envelope (CE) event in the inner binary for its dynamical stability.

Consider a stable hierarchical triple system, composed of a tight inner binary with component masses $m_{\rm 1}$ $\ge$ $m_{\rm 2}$ that is orbited by a tertiary of mass $m_{\rm 3}$. All stars are initially assumed to be on the main-sequence and, for the remainder of this section, we assume $m_{\rm 1}$ $\ge$ $m_{\rm 2}$ $>$ $m_{\rm 3}$, but relax this assumption later on.  The inner and outer orbital semi-major axes are initially denoted by $a_{\rm in}$ and $a_{\rm out}$, respectively.  We assume both orbits, the inner as well as the outer, to be initially circular and in the same plane.  In principle, we expect that this should minimize chaotic effects during the mass transfer process, facilitating more stable mass transfer and ultimately maximizing the amount of mass transferred (and therefore facilitating the production of blue stragglers).  These assumptions are supported by the population of observed low-mass triples, since co-planar orbits are typically observed when the outer orbit is small (i.e., $\lesssim$ 10$^{4}$ R$_{\rm \odot}$) \citep[e.g.][]{kouwenhoven08,2010yCat..73890925T,2018ApJ...854...44M,tokovinin18}.  

Now, we assume that the inner binary experiences a common envelope (CE) event, due to one of its two companions (m$_{\rm 1}$) expanding upon evolving off the main-sequence.  The CE expands adiabatically beyond the orbit of the inner binary until it reaches the outer triple companion.  We consider two extreme cases:  (1) the outer tertiary accretes none of the expanding mass, which we assume escapes via a fast spherically symmetric wind; and (2) the tertiary accretes all of the mass via the inner L1 Lagrangian point in the outer orbit.  We emphasize that the first extreme case ignores changes to the outer orbital period which could occur due to asymmetric mass loss and/or friction, but such effects would need to be quantified via detailed numerical simulations which is beyond the scope of this work.  The second extreme case is justified by the fact that the Bondi radius in the tertiary is much larger than the stellar radius for small gas sound speeds on the order of $c_{\rm s} \lesssim$ 10 km s$^{-1}$ \citep{portegieszwart19}.

\begin{figure}
\includegraphics[width=\columnwidth]{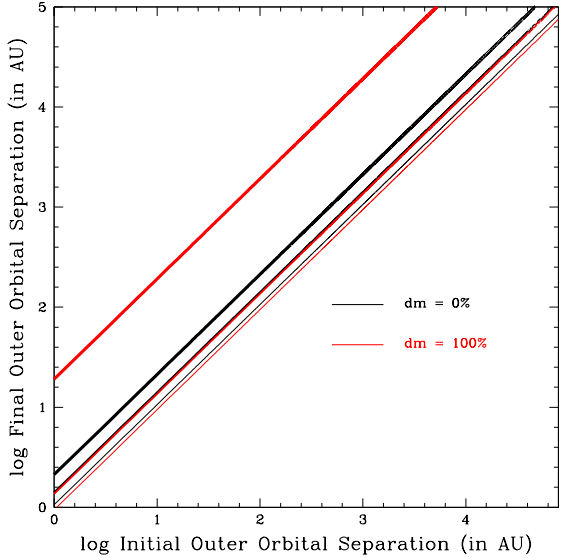}
\caption{The initial and final outer orbital separations for the outer tertiary orbit due to the ejection of a common envelope in the inner binary.  The black lines assume no mass accretion by the tertiary whereas the red lines assume 100\% mass accretion.  The line widths correspond to different total fractions of the total inner binary mass that is lost via the ejection of its common envelope.  We adopt 10, 50 and 90\% for the thin, medium and thick lines, respectively.
\label{fig:fig1}}
\end{figure}

Figure~\ref{fig:fig1} shows the initial and final outer orbital separations for the aforementioned competing assumptions.  The black lines assume no mass accretion by the tertiary (i.e., non-conservative mass transfer, or a spherically symmetric adiabatic wind with no accretion from the tertiary) whereas the red lines assume 100\% mass accretion (i.e., conservative mass transfer, with all of the mass lost from the inner binary accreted by the tertiary companion).  To compute the final orbits for the non-conservative mass transfer case, we assume:
\begin{equation}
\label{eqn:noncons}
\frac{a_{\rm f}}{a_{\rm i}} = \frac{m_{\rm 1} + m_{\rm 2} + m_{\rm 3}}{m_{\rm 1} + m_{\rm 2} + m_{\rm 3} - dm},
\end{equation}
where $a_{\rm f}$ and $a_{\rm i}$ are, respectively, the final and initial orbital separations for the outer tertiary orbit and $dm$ is the fraction of mass lost from the inner binary.  To compute the final orbits for the conservative mass transfer case, we assume:
\begin{equation}
\label{eqn:cons}
\frac{a_{\rm f}}{a_{\rm i}} = \left[ \frac{(m_{\rm 1} + m_{\rm 2})m_{\rm 3}}{(m_{\rm 1} + m_{\rm 2} - dm)(m_{\rm 3} + dm)} \right]^2,
\end{equation}
The line widths correspond to different total fractions of the total inner binary mass that is lost via the ejection of its common envelope.  We adopt 10, 50 and 90\% for the thin, medium and thick lines, respectively, and emphasize that 90\% mass loss would be extreme and should be regarded as a strict upper limit.  Note that we have not considered accretion onto $m_{\rm 2}$ in this scenario.  This is possible, but in general not expected \citep{iben87,ivanova13}.

We caution that we assume that, in the non-conservative mass transfer case, the outer orbit will always remain bound.  However, this is only correct if the timescale for ejecting the CE is much longer than the outer orbital period.  In the opposite regime, the CE mass loss is effectively instantaneous and, in this limit, it is possible to unbind the tertiary orbit \citep{iben99}.  The assumption of adiabatic mass loss is justified since we expect, for the compact triples considered here, that the timescale for ejecting the CE will greatly exceed the outer orbital period.  With that said, however, the CE ejection timescale is poorly constrained \citep[e.g.][]{glanz18}, although \citet{erez19} recently introduced a method to use prior CE evolution in observed wide triples to constrain the timescale for CE ejection.

The take-away message from Figure~\ref{fig:fig1} is that, for most of the relevant parameter space, the outer orbit expands due to the mass transfer-induced or (CE-induced) ejection event.  It should therefore expand again if a second mass transfer- (or CE-) induced ejection event occurs in the inner binary, producing a final inner binary composed of two compact objects.  If more mass is accreted by the outer triple companion, then the final outer orbit is wider.  It follows that, given that we expect the inner orbit to become more compact by losing orbital energy and angular momentum to help power the expansion of the CE \citep{webbink84}, the ratio $a_{\rm out}$/$a_{\rm in}$ should increase.  Thus, this simple exercise suggests that, if a given triple was initially dynamically stable, then it should also typically remain stable over the course of any CE event in the inner binary.  In fact, it is likely to become even more dynamically stable as a result of binary evolution and mass loss in the inner binary.  We caution, however, that we have invoked simplifying assumptions that should be approximately valid to first order but would need to be quantified via hydrodynamics simulations, such as spherically symmetric mass loss and no friction acting on the outer tertiary orbit.


\subsection{The Post-Stellar and Binary Evolution Triple} \label{post}

%

We now consider the implications of Roche lobe overflow in the outer tertiary companion.  We focus our attention on an inner binary composed of two WDs for illustrative purposes (but relax this assumption later on), and the possible formation of a circumbinary disk orbiting around it.   

Following from the scenario discussed in the previous section, once the inner binary has evolved to be composed of two compact objects in a relatively tight orbital configuration, we have $m_{\rm 3} >$ $m_{\rm 1}$, $m_{\rm 2}$.  The outer orbit is assumed to be 
sufficiently small that the tertiary star will overfill its Roche lobe at some point over the course of its evolution, 
and transfer mass to the inner binary while remaining dynamically stable. We constrain the inner and outer orbits by requiring the triple
system to be dynamically stable, for which we use as a guide Equation 1 in
\cite{1999ASIC..522..385M} and the calculation in Section 5.2 in \citet{tokovinin18}.  Specifically, we adopt the criterion from \citet{mardling01} to determine the minimum value of the ratio $a_{\rm out}$/$a_{\rm in}$ $\ge$ $R_{\rm 0}$ for which the outer tertiary orbit should be dynamically stable \citep{tokovinin18}:
\begin{equation}
\label{eqn:stable}
R_{\rm 0} = 2.8(1 + q_{\rm out})^{1/15}(1 + e_{\rm out})^{0.4}(1 - e_{\rm out})^{-1.2},
\end{equation}
where $q_{\rm out}$ and $e_{\rm out}$ are, respectively, the mass ratio and eccentricity of the outer tertiary orbit.  We assume co-planar orbits and $m_{\rm 3} =$ 1 M$_{\odot}$ for the outer tertiary, which corresponds to a rough lower limit for $R_{\rm 0}$ since the majority of the allowed parameter space gives more massive tertiaries.  We assume $e_{\rm out} =$ 0 and $q_{\rm out} =$ $m_{\rm 3}$/($m_{\rm 1} +$ $m_{\rm 2}$), where the component masses of the inner binary (i.e., $m_{\rm 1}$ and $m_{\rm 2}$) are either taken to be the assumed initial mass for MS stars or the WD mass at the time of formation as calculated by \texttt{SeBa}.

We assume that, while transferring mass, the accretion
stream from the outer tertiary gathers around the inner binary at the circularization radius
$a_{\rm c}$, and forms a circumbinary disk
\citep{2002apa..book.....F}.  Using conservation of angular momentum,
we equate the specific angular momentum of the accreted mass at the
inner Lagrangian point of the (outer) donor star to the final specific
angular momentum of the accretion stream at the circularization radius
about the inner binary.  This gives:
\begin{equation}
\label{eqn:specangmom1}
v_{\rm orb,3}a_{\rm out}(1 - R_{\rm L}) = v_{\rm orb,c}a_{\rm c},
\end{equation}
where $R_{\rm L}$ is the relative radius of the Roche lobe of the outer
tertiary companion in terms of the orbital separation, $a_{\rm c}$ is the semi-major axis of the orbit
about the inner binary corresponding to the circularization radius and
$v_{\rm orb,c}$ is the orbital velocity at $a_{\rm c}$.  The distance
from the centre of mass corresponding to the tertiary defined by the
Roche lobe is given by Equation 2 in \cite{1983ApJ...268..368E}.
Combining Equation 2 in \citet{1983ApJ...268..368E} (with mass ratio q
$=$ $m_{\rm 3}$/($m_{\rm 1} +$ $m_{\rm 2}$)) with
Equation~\ref{eqn:specangmom1}, we solve for the circularization
radius as a function of $a_{\rm out}$ and the assumed stellar masses:
\begin{equation}
\label{eqn:ac}
a_{\rm c} = a_{\rm out}(1 - R_{\rm L})^2.
\end{equation}
In order for a circumbinary disk to form around the inner binary, we
require that $a_{\rm in} < a_{\rm c}$.

\begin{figure}
\includegraphics[width=\columnwidth]{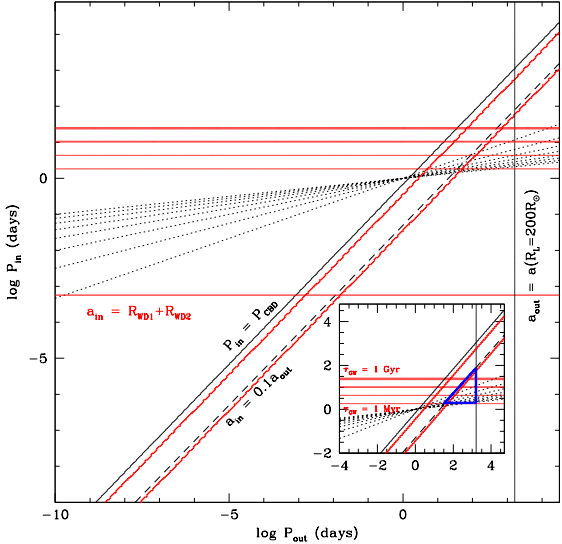}
\caption{Parameter space in the $P_{\rm out}$-$P_{\rm in}$-plane
  allowed for the hypothetical outer tertiary orbit 
  before Roche-lobe overflow.  
Please refer to the text for more details.
\label{fig:fig2}}
\end{figure}

Figure~\ref{fig:fig2} shows the parameter space in the $P_{\rm out}$-$P_{\rm in}$-plane for a hypothetical WD-WD inner binary, where $P_{\rm in}$ and $P_{\rm out}$ denote, respectively, the periods of the inner and outer orbits.  Here we adopt 
initial component masses of $m_{\rm 1} = 0.4$ M$_{\rm \odot}$
and $m_{\rm 2} = 0.55$ M$_{\rm \odot}$ for the inner WD binary components,
and $m_{\rm 3} = 1.4 $M$_{\rm \odot}$ for the outer tertiary, computed 
according to our assumed mass ratio (with our fiducial case
corresponding to $q$ $=$ 0.7).  Although these specific
parameters are chosen somewhat arbitrarily, they appear to naturally result in a system with
parameters similar to those one might expect for pre-merger tight WD-WD binaries.  We compare the circularization
radius to the semi-major axis of the inner binary, for which we
require $a_{\rm c} > a_{\rm in}$, after folding in all constraints
from the requirements for dynamical stability, 
and the assumption of an outer tertiary
that is Roche lobe-filling (see \citet{2014MNRAS.438.1909D} for more
details).  Specifically, the diagonal dashed 
black line shows a rough criterion for dynamical stability in the
triple, approximately following \citet{1999ASIC..522..385M} (i.e.,
$a_{\rm in} \lesssim$ 0.1$a_{\rm out}$ is required for long-term dynamical
stability in equal-mass co-planar triples).
The vertical solid black line shows the maximum
outer orbital period $P_{\rm out}$ for which the outer tertiary
companion is Roche lobe-filling, assuming a stellar radius of $R_{\rm
  3} =$ 200 R$_{\odot}$ (which roughly corresponds to the maximum stellar
radius reached on the AGB for the range of tertiary masses of interest
to us).  The diagonal dotted black lines show orbital resonances (from 1:3 up to 1:10) between 
the inner and outer orbits of the triple.  The horizontal solid red lines show the timescale 
for merger due to GW emission, at the indicated orbital period.  From thinnest to thickest lines, we 
show $\tau_{\rm GW} =$ 1 Myr, 10 Myr, 100 Myr, 1 Gyr.  Finally, the diagonal solid black line shows the location 
of the 1:1 orbital resonance between the inner binary and the CBD (at the circularization radius) for zero eccentricity.  The top diagonal solid red line shows the location of the resonance 
assuming that the inner binary semi-major axis is half that of the CBD.  
The bottom diagonal solid red line assumes an even more compact inner orbit by a factor of 0.1.

The initial orbital period of the outer tertiary companion is surprisingly well constrained, before any mass transfer from the tertiary onto the inner binary.  This is because the system must live in one of the triangles (shown in the inset) bounded by the diagonal dashed black line (otherwise the triple would not be dynamically stable), the vertical solid black line (otherwise the outer tertiary would not be Roche lobe filling, to first order) and one of the horizontal solid red lines (otherwise the inner binary would merge due to GW emission too quickly, and would hence not survive for sufficiently long to have a non-negligible probability of being observed).  We have indicated one example (i.e., adopting a timescale for merger due to GW emission of 1 Myr) of this triangular region by highlighting it using thick blue lines, as shown in the lower right inset.\footnote{The corresponding ranges for the triangles are decided by the chosen timescale for merger due to GW emission (i.e., the horizontal solid red lines).  For example, if we choose $\tau_{\rm GW} =$ 1 Myr, then the corresponding ranges are approximately 1.5 $<$ log $P_{\rm out} <$ 3.2 on the x-axis and 0.3 $<$ log $P_{\rm in} < 1.7$ on the y-axis.}  As discussed in more detail in the next section, these initial orbital periods should be relatively representative of the final orbital periods, but this depends on how conservative the mass transfer process is from the tertiary onto the inner binary, which will require detailed hydrodynamics simulations to quantify.  As described in the preceding section, it also depends on the timescale for CE ejection and how this compares to the outer tertiary orbital period, with our assumption of mass loss being due to a fast spherically symmetric wind corresponding to the case where the timescale for ejecting the CE is much longer than the outer tertiary orbital period.  The preliminary results of \citet{portegieszwart19} suggest non-conservative mass transfer, which would leave the final outer tertiary orbital period within a factor of a few of the initial orbital period.


\subsection{Binary Evolution Models} \label{binevmodels}

In this section, we incorporate binary evolution models performed using the \texttt{SeBa} code \citep{portegieszwart96,toonen12}, to more robustly constrain the expected frequency of tight inner binaries with outer triple companions that will evolve to become simultaneously Roche lobe over-filling and dynamically stable.

The models used in this paper are performed using the \texttt{SeBa} code for stellar and binary evolution (see \citet{portegieszwart96} and \citet{toonen12} for more details) and comprise the default models for WD populations with \texttt{SeBa} \citep[see e.g.][]{Too17}. We begin by simulating the evolution of 250,000 isolated MS-MS binaries and focus on those that evolve to contain white dwarf components. We take into account processes such as stellar winds, mass transfer and tides. For the initial binaries we have assumed that 1) the masses of the primary stars follow the initial mass function of \citep{Kro93} between 0.1-100$M_{\odot}$; 2) the mass ratios are drawn from a uniform distribution between 0 and 1; 3) the logarithm of the orbital separations follow a uniform distribution up to $10^6R_{\odot}$ \citep{Abt83, Moe17}; 4) the eccentricities are drawn from a thermal distribution \citep{Heg75} between 0 and 1; 5) the metallicity is solar. We only include binaries that are initially detached.  Regarding point 1, we adopt two sets of MS-MS models.  The first assumes initial primary masses in the range 0.1 - 100 M$_{\odot}$, whereas the second considers only initial primary star masses in the range 0.95 - 10 M$_{\odot}$.  The latter range corresponds to roughly 10\% of the full range, and it is important to note that stars with masses smaller than this tend to be the majority.  With that said, most of these lower primary masses are unlikely to evolve off the MS within the age of the Universe, and hence are mainly of interest for our MS-MS case (see Figure~\ref{fig:silvia4}).


\citet{Too14} showed that the main sources for differences between synthesized populations from different codes is due to the choice of input physics and initial conditions, in particular for the physics of unstable mass transfer i.e. common-envelope evolution \citep[see][for a review]{Iva13}. For this reason, when evolving the inner orbit due to binary evolution, we adopt the two standard models adopted in \texttt{SeBa} which are fully described in \cite{toonen12} (that is, their model $\alpha\alpha$ and $\gamma\alpha$). In short, in our fiducial case the common-envelope phase is modeled based on the conservation of energy comprising the orbital energy and the binding energy of the primary's envelope \citep{Pac76,webbink84,Liv88}. For the second set of models we adopt an alternative prescription for the common-envelope phase where the binary does not contain a compact object or the CE is not triggered by a tidal instability, as proposed by \citet{Nel00} to explain the formation of double white dwarfs. In this alternative method, the common-envelope is based on a balance of angular momentum.  

For each binary, we place a third star on a wide orbit centred on the centre of mass of the binary (hereafter inner binary), assuming a mass of 1 M$_{\odot}$ for the tertiary companion.  We compute the range of outer tertiary orbital periods that would be able to host a CBD around the inner binary by simultaneously satisfying the constraints to be both Roche lobe over-filling and dynamically stable.  
This produces a final sample of triples with inner binaries that have survived their internal evolution, which is what we use to compute the fraction of systems expected to be able to host a CBD around the inner binary (see below and Equation~\ref{eqn:ac}). We implicitly assume that three-body dynamics \citep[e.g. Kozai-Lidov cycles][]{Koz62, Lid62} have a marginal effect on the orbital evolution of the triple, as triples with rather compact outer orbits tend to be orientated coplanar \citep{tokovinin18}.\footnote{We note that such three-body dynamics of initially inclined orbits would tend to reduce the orbital separation of the inner binary, which could make the inner binary components interact sooner than computed by \texttt{SeBa}.  Moreover, coplanar hierarchical triples can actually be interacting.  For example, octupole-order terms in the secular equations of motion could still drive high inner binary eccentricities if the outer orbit is also eccentric (see \citet{li14} for more details.)}  To determine the fraction of triples that can remain dynamically stable at least until the formation of the CBD, we apply Equation~\ref{eqn:stable} to the 
inner binary at the moment of its maximum orbital separation during its evolution.

\begin{figure}
\includegraphics[width=\columnwidth]{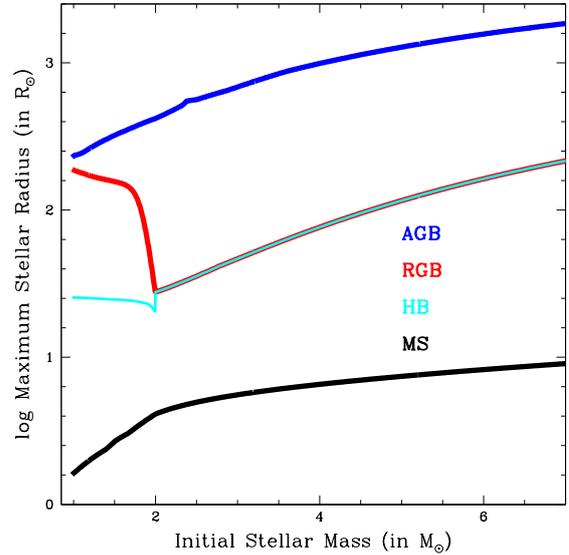}
\caption{The maximum stellar radius (in R$_{\odot}$) reached as a function of stellar mass (in M$_{\odot}$) in \texttt{SeBa}, corresponding to different phases of stellar evolution.  The dark blue, red, cyan and black lines correspond to, respectively, the maximum stellar radius reached on the asymptotic giant branch (AGB), red giant branch (RGB), horizontal branch (HB) and main-sequence (MS).  
\label{fig:silvia1}}
\end{figure}

\begin{figure}
\includegraphics[width=\columnwidth]{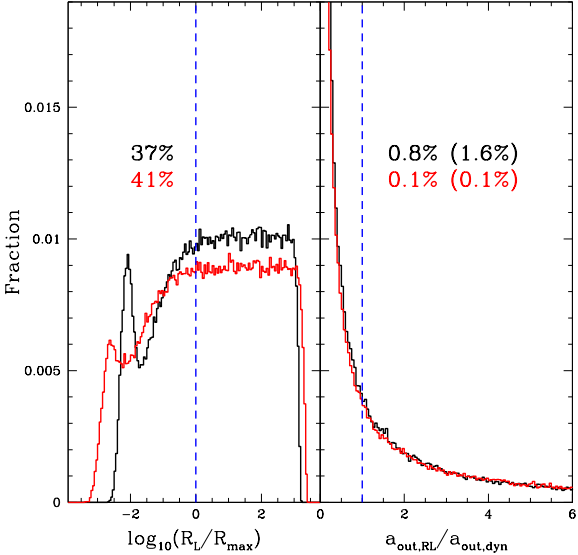}
\caption{The fraction of triple systems hosting an inner compact MS-MS binary and an outer main-sequence tertiary companion that will overflow its Roche lobe at some point over the course of its evolution.  The left panel indicates the fraction of systems that will overflow their Roche lobes at some point over the course of their evolution (see Figure~\ref{fig:silvia1} and the text for more details), whereas the right panel indicates those systems for which the outer tertiary will overfill its Roche lobe while simultaneously maintaining dynamical stability.  The black histograms correspond to our initial MS-MS binary population assuming initial primary star masses in the range 0.95 - 10 M$_{\odot}$, whereas the red histograms assume initial primary masses in the range 0.1 - 100 M$_{\odot}$.  The former range corresponds to roughly 10\% of the full range, since stars with masses smaller than this tend to be the majority.  The main point to take away is that the distributions are largely unaffected by the chosen mass ranges for the inner binary components.  
In the left panel, the vertical dashed blue line demarcates the critical ratio below which the criterion for Roche lobe overflow is satisfied, and the provided percentage indicates this fraction.  In the right panel, we show the fraction of simulated systems satisfying 0.95 $<$ $a_{\rm out,RL}$/$a_{\rm out,dyn} <$ 1.05.  In parantheses, we indicate those systems satisfying 0.90 $<$ $a_{\rm out,RL}$/$a_{\rm out,dyn} <$ 1.10.
\label{fig:silvia4}}
\end{figure}

To determine the fraction of systems for which the tertiary will fill its Roche lobe at some point during its evolution, we check the maximum stellar radius reached as a function of stellar mass, at different phases of stellar evolution as given by \texttt{SeBa} (see Figure~\ref{fig:silvia1}). We adopt the maximum stellar radius for a 1 M$_{\odot}$ star when on the asymptotic giant branch (AGB) in making Figures~\ref{fig:silvia4}, ~\ref{fig:silvia3} and~\ref{fig:silvia2} (see below).  As we will show, this should correspond to a rough lower limit for the outer orbital separation at the time of mass transfer due to the low assumed mass for the tertiary, but a rough upper limit for the outer orbital separation in the sense that our calculations assume the maximum stellar radius possible at the time of mass transfer. 

Figures~\ref{fig:silvia4}, ~\ref{fig:silvia3} and~\ref{fig:silvia2} show the fraction of our triple systems (relative to the total population of triples that \texttt{SeBa} computes should survive their internal evolution) that are capable of hosting, respectively, an inner compact MS-MS, MS-WD and WD-WD binary and an outer MS tertiary companion with a mass of 1 M$_{\odot}$ that will overflow its Roche lobe at some point over the course of its evolution.  In Figures~\ref{fig:silvia3} and~\ref{fig:silvia2}, the black lines correspond to our fiducial binary evolution models, whereas the red lines show our results for the more efficient CE prescriptions.  The left panel shows those systems that will at some point overflow their Roche lobes (see Figure~\ref{fig:silvia1}), whereas the right panel shows those systems for which the outer tertiary will \textit{both} overfill its Roche lobe while \textit{also} simultaneously maintaining dynamical stability.  In the left panel, it is those systems with log ($R_{\rm L}$/$R_{\rm max}$) $<$ 0 that satisfy our criterion for Roche lobe overflow, with the total number of such systems to the left of the vertical dashed lines corresponding to the numerator in our calculation for the expected fraction of systems.  In the right panel, $a_{\rm out,RL}$ denotes the maximum outer orbital separation that can accommodate Roche lobe overflow and $a_{\rm out,dyn}$ denotes the minimum outer orbital separation for dynamical stability (over the entire evolution of the inner binary).  Our criterion is satisfied provided $a_{\rm out,RL}$/$a_{\rm out,dyn}$ $\gtrsim$ 1 since only here is the outer orbit both tight enough to be Roche lobe filling and wide enough to maintain dynamical stability.  The allowed range about this constraint is narrow, since the outer orbit cannot become very wide for a 1 M$_{\odot}$ tertiary to be Roche lobe filling and the inner binary tends to expand during its internal evolution.  Said another way, it is not always the case that dynamical stability also implies $a_{\rm in}$ $<$ $a_{\rm c}$.  For example, as occurs most frequently for very wide inner binaries, the outer tertiary must also be on a very wide orbit in order to guarantee dynamical stability, but this  lowers the probability that the tertiary will become Roche lobe overfilling.

As is clear from the indicated percentages in Figures~\ref{fig:silvia4}, ~\ref{fig:silvia3} and~\ref{fig:silvia2}, for our assumed initial conditions, the expected frequency of triple star systems expected to simultaneously satisfy both constraints that the outer tertiary must become Roche lobe-filling while also maintaining dynamical stability (and hence being able to accommodate an inner binary orbital separation smaller than the predicted circularization radius for a CBD) is of order a few percent, specifically $\sim$ 1\%, $\sim$ 1-5\% and 5-10\% for, respectively, MS-MS, WD-MS and WD-WD binaries.  
Given that we have adopted a 1 M$_{\odot}$ tertiary star and this corresponds to a lower limit for our criteria, our results suggest that of order $\gtrsim$ 10\% of all primordial MS triples should evolve to be able to accommodate a CBD around the inner binary, given a reasonable population of initial triples motivated by the available empirical data \citep{tokovinin18}.

\begin{figure}
\includegraphics[width=\columnwidth]{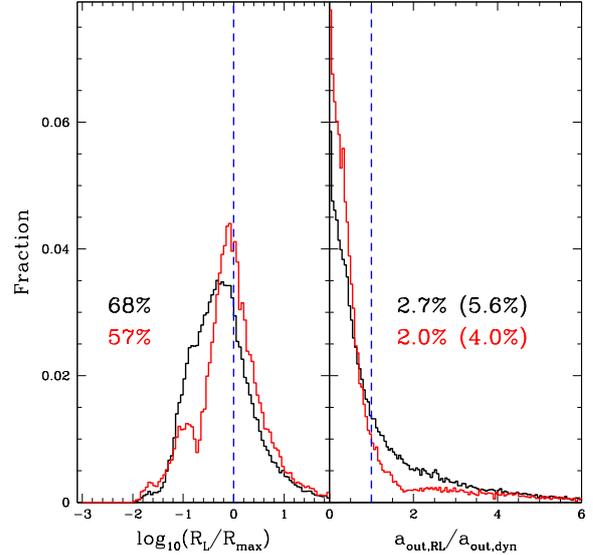}
\caption{The same as Figure~\ref{fig:silvia4}, but showing instead the fraction of triple systems hosting an inner compact WD-MS binary.  The black and red histograms show, respectively, our fiducial set of models and the set assuming our modified CE prescription. 
\label{fig:silvia3}}
\end{figure}

\begin{figure}
\includegraphics[width=\columnwidth]{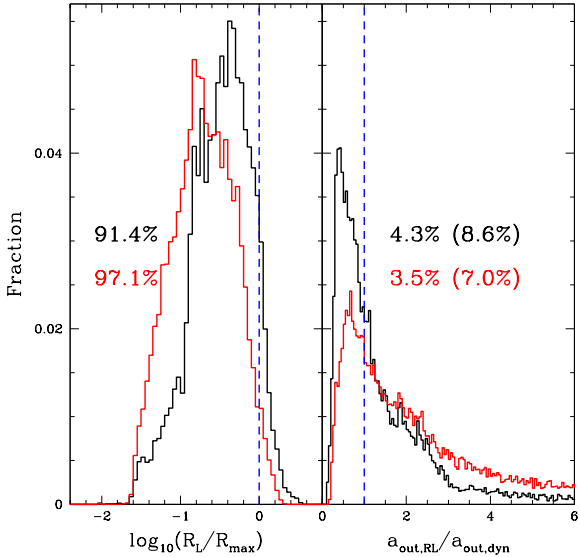}
\caption{The same as Figure~\ref{fig:silvia3}, but showing instead the fraction of triple systems hosting an inner compact WD-WD binary.  
\label{fig:silvia2}}
\end{figure}

Based on Figure~\ref{fig:silvia1}, more massive outer tertiary companions will overflow their Roche lobes at larger outer orbital separations than shown in Figures~\ref{fig:silvia4}, ~\ref{fig:silvia3} and~\ref{fig:silvia2}.  Thus, we expect this to contribute to an \textit{increase} in the predicted fraction of stable hierarchical triples with an outer tertiary companion that will eventually overfill its Roche lobe, for more massive outer tertiary companions than considered in Figures~\ref{fig:silvia4}, ~\ref{fig:silvia3} and~\ref{fig:silvia2}.  This is simply because the requirement for smaller outer orbital separations for low-mass tertiaries implies a smaller volume of available parameter space for stable mass transfer onto the inner binary from a CBD.  \textit{Therefore, since our adopted mass for the outer tertiary companion in Figures~\ref{fig:silvia4}, ~\ref{fig:silvia3} and~\ref{fig:silvia2} is low, our results can be regarded as lower limits relative to any more complete and/or realistic sample of triple star systems that accounts for the full range of allowed tertiary masses.}  The only caveat to this is that larger tertiary masses tend to reduce the parameter space for dynamical stability via Equation~\ref{eqn:stable}, since $R_{\rm 0}$ increases with increasing $q_{\rm out}$.  However, the increase in parameter space due to more massive tertiaries being able to overfill their Roche lobes at larger outer orbital separations is much more significant than the aforementioned decrease due to dynamical instability, since the dependence of $R_{\rm 0}$ on $q_{\rm out}$ is quite weak.

\begin{figure}
\includegraphics[width=\columnwidth]{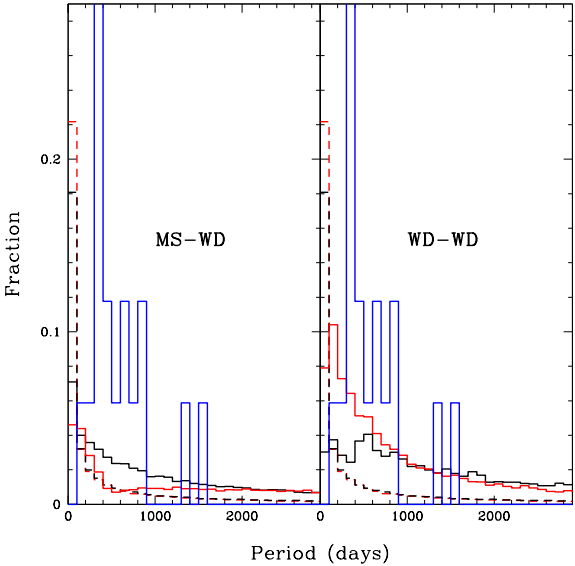}
\caption{The predicted outer orbital periods of triple systems hosting an inner compact MS-WD (left panel) or WD-WD (right panel) binary and an outer 1 M$_{\odot}$ main-sequence tertiary companion that will overflow its Roche lobe at some point over the course of its evolution while simultaneously maintaining dynamical stability.  As before, the black and red solid histograms show our results for, respectively, our fiducial case (black) and those binary evolution models adopting a more efficient CE prescription (red).  Note that the more efficient CE prescription models systematically predict more compact inner binaries, and hence more compact outer tertiary orbits.  
The black dashed histograms correspond to our initial MS-MS binary population assuming initial primary star masses in the range 0.95 - 10 M$_{\odot}$, whereas the red dashed histograms assume initial primary masses in the range 0.1 - 100 M$_{\odot}$.  The former range corresponds to roughly 10\% of the full range, and stars with masses smaller than this tend to be the majority.  The main point to take away is that the distributions are largely unaffected by the chosen mass ranges for the inner binary components.  
The blue histograms show a comparison to the observed BS binaries in the field presented in Table 5 of \citet{carney05}.
\label{fig:silvia5}}
\end{figure}

We point out that the results shown in Figures~~\ref{fig:silvia4}, ~\ref{fig:silvia3} and~\ref{fig:silvia2} for the predicted outer orbital periods of the tertiary companion are in good agreement with what is shown in Figures~\ref{fig:fig1} and Figure~\ref{fig:fig2} using analytic calculations.  This is illustrated in Figure~\ref{fig:silvia5}, which shows the predicted distributions of outer tertiary orbital periods assuming the most compact state allowed from dynamical stability while being Roche lobe filling (and satisfying the condition $a_{\rm in} < a_{\rm c}$) for both MS-WD and WD-WD inner binaries, as well as for our assumed initial population of MS-MS binaries.  These represent the periods before the outer tertiary evolves off the main-sequence and turns into, most likely, another WD.  But they should be very similar to the final outer orbital periods (i.e., post mass-transfer from the outer tertiary and the formation of any CBD).  This is because, by conservation of orbital energy and angular momentum, the outer orbital separations should not change by more than a factor of a few due to mass loss from the inner binary (assuming spherically symmetric mass loss via a fast wind), bearing in mind that the outer orbit cannot expand much before the tertiary would no longer be Roche lobe filling.  Figure~\ref{fig:silvia5} predicts that of order half of all tertiary orbital periods capable of hosting a CBD should lie in the range of a few times $\sim$ 100-1000 days. We caution, however, that the final tertiary orbital period post-mass transfer depends on how conservative the mass transfer process is.  The preliminary simulations of \citet{portegieszwart19} suggest that the mass transfer can be highly non-conservative, which supports our previous assumptions.  Nevertheless, understanding to what extent mass is conserved will require a more detailed exploration of the relevant parameter space than considered in \citet{portegieszwart19}.

Finally, in Figure~\ref{fig:silvia5}, we overplot the field BS sample from \citet{carney05}, specifically the periods shown in their Table 5.  We note that these authors conclude that all orbital solutions for this BS sample are consistent with all unseen companions in their long-period low-eccentricity binaries being WDs viewed at random orbital orientations. This is based on their evaluation of the minimum binary companion masses to their observed BSs.  

Next, we compare our sample of predicted triples with the BS sample from \citet{carney05} to assess the degree to which they are similar and, in so doing, attempt to address whether or not the triple mechanism proposed here could have actually produced the observed BS binaries.  We caution, however, that the \citet{carney05} sample suffers from low-number statistics, and the comparison should be performed both carefully and critically.  A Kolmogorov-Smirnov (KS) test comparing the BS distribution from \citet{carney05} to the outer tertiary orbital periods of those MS-MS inner binaries able to accommodate a CBD suggests that the two distributions are significantly different.  However, this is not so surprising for three reasons.  First, the predicted sample includes many wide inner binaries, with even wider outer tertiaries that can still satisfy the constraints for dynamical stability and Roche lobe overflow.  Given that significant energy and angular momentum would need to be removed from the inner orbit, the wide inner binaries of these systems are unlikely to merge, and should likely be removed from the sample comparison, since they are unlikely to produce a single BS companion in a long-period binary with a WD companion.  Second, there is an observational bias in the observed sample, namely a maximum detectable orbital period.  This limit is imposed by the time baseline of the radial velocity survey used to identify the binary orbital periods, and constrains the observed sample to have periods of a few thousand days or less.  Third, we have not corrected the outer orbit due to mass transfer onto the inner binary, which is needed for a proper comparison between our predicted long-period BS-WD binaries and the observed sample of BS binaries in \citet{carney05}.  

To see if taking into account the first and third effects could improve the agreement, we re-perform our KS test after shifting the entire predicted orbital period distribution by an arbitrary constant.  The results of this exercise suggest that, if the outer orbit shrinks as a consequence of the mass transfer and/or merger in the inner binary, then the agreement could improve significantly.  In reality, the outer orbit is unlikely to shrink, since both mass transfer and mass loss should tend to widen the orbit.  However, again, this improved agreement is most likely due to reducing the periods of those wide inner binaries with even wider outer tertiaries in our predicted sample for which we do not expect the inner binary to merge.  

To see if taking into account the second effect could improve the agreement of our KS test, we try truncating the theoretical samples at an outer orbital period of 2000 days.  This does in fact improve the agreement significantly, supporting yet again the idea that the longest period orbits in our theoretical samples should be removed from the samples for a more proper comparison.  We also try removing the shortest period systems from the observed sample of \citet{carney05}, which would also improve the agreement.  Hence, if these few systems were formed from an alternative channel then considered here, in which case they should not be included in the comparison, this too would improve the agreement with our predicted distribution.  We note that, for example, these short-period systems could form via an inner MS-WD binary accreting from a CBD in a triple, with the outer triple companion having remained thus far undetectable (e.g., due to a long orbital period that exceeds the time baseline of the radial velocity study performed by \citet{carney05}).

Independent of the above KS test, our results are still broadly consistent with the population of BSs shown in \citet{carney05}, since they can explain why these systems tend to have orbital periods in the range of a few times $\sim$ 100-1000 days, which is required for our triples to become Roche lobe-filling while maintaining dynamical stability. \citet{carney05} note, however, that a previous study by \citet{rappaport95} showed that these systems are consistent with having formed from stable mass-transfer in a progenitor MS-MS binary when one of the objects evolves to overfill its Roche lobe.  This would not, however, explain the reason for the small range of periods derived by \citet{carney05} whereas the triple mechanism considered in this paper could.

\section{Discussion} \label{sect:discussion}

In this paper, we propose a formation scenario for equal-mass tight binaries in triples hosting COs, with a focus on WDs.  The proposed scenario involves mass transfer from an evolved outer tertiary companion. At least some of this mass is accreted by the inner binary via a circumbinary disk.  We focus on those evolutionary channels that initiate accretion onto either one or more MS stars or one or more WDs in the inner binary via an outer tertiary companion, producing one or more types of stellar exotica (blue stragglers, Type Ia supernovae, cataclysmic variables, and so on) at some point over the course of their evolution.   Our scenario makes several predictions for the final observed properties of the various evolutionary pathways for the hypothetical initial populations of MS triples considered here, which we review and discuss in this section. We list our primary conclusions and present our predictions, along with their implications for the production of various types of stellar exotica.  We begin with triples containing MS stars and/or WDs in the inner binary, since this is our focus in this paper, but then go on to extend the results presented in this paper to consider triples containing neutron stars and/or black holes in the inner binary.

\subsection{Triples with inner binaries containing MS stars and/or WDs}

Below, we list our primarily conclusions and predictions for the observed properties of the considered evolutionary channels for triples with MS stars and/or WDs in the inner binary, in addition to their implications for the production of various types of stellar exotica.

\begin{itemize}

\item 
As shown in Section~\ref{prog}, any common envelope event in the inner binary will most likely preserve the dynamical stability of any outer tertiary companions.  Given our assumptions (i.e., mass loss via a fast spherically symmetric wind and no friction acting on the tertiary orbit), this is roughly independent of the amount of mass lost from the inner binary, and/or the amount of mass accreted by the outer tertiary companion.  


\item Below, we individually consider each possible evolutionary pathway for our hypothetical triple star systems, as decided by stellar evolution and hence the relative masses between the three components of our triples \textit{before} any mass transfer from the tertiary companion has occurred.  We begin with the assumption that the inner binary contains no WDs when the outer tertiary overfills its Roche lobe, and proceed by sequentially replacing each MS star in the inner binary with a WD.

\subitem i) \textit{Accretion onto a MS-MS inner binary (i.e., $m_{\rm 3}$ $>$ $m_{\rm 2}$ $>$ $m_{\rm 1}$)}

This scenario was considered in detail in \citet{portegieszwart19}, and quantified via hydrodynamics simulations.  The authors showed that, if a stable CBD forms around the inner binary, then at least one BS should form, producing at least one BS in a short-period binary system, with an outer tertiary companion that is most likely a WD.  \citet{portegieszwart19} further argued that the production of compact twin BS binaries (with mass ratios close to unity) could in fact be the dominant channel or evolutionary pathway.  We note briefly here a possible connection to contact binaries, called W UMa systems, many of which are know to host BSs \citep{rucinski10}.  The approximate distribution of predicted orbital periods for this mechanism are shown via the dashed black histograms in Figure~\ref{fig:silvia5}.  

Double BS binaries will eventually evolve into double WD binaries (unless they merge prior to this; see below).  By construction, the tertiary is the most massive companion in this scenario, such that the two MS stars in the inner binary tend to have low masses.  Consequently, from single star evolution theory alone, we expect most MS stars in the inner binary to eventually evolve into CO WDs.  The outer tertiary companion, however, would be more likely to leave behind either a massive CO WD or an ONe WD (with the maximum mass for a CO WD from single star evolution thought to lie around 1.1 M$_{\odot}$ \citep[e.g.][]{hollands20,portegieszwart96,toonen12}). Hence, of the three possibilities considered here, this scenario predicts the largest mass ratios, where $q$ $=$ $m_{\rm 3}$/($m_{\rm 1} +$ $m_{\rm 2}$), independent of the evolutionary status of the components (i.e., WDs or MS stars).

If the inner binary pair merges while still on the MS, it will turn into an over-massive BS.  If in a star cluster, we expect the total BS mass to exceed twice the mass corresponding to the cluster main-sequence turn-off. The merged BS should have a high rotation rate (at least initially), and be in a long-period binary with a WD companion.  Hence, as touched upon in \citet{leigh11} (see Figure 3 and the corresponding discussion), the tertiary formation route with mass transfer can in principle form most of the BS systems observed in old open clusters (e.g., NGC 188, M67) and the field \citep[e.g.][]{mathieu09,geller15,peterson84}, including both long- and short-period BS systems.  As pointed out in \citet{leigh11}, this is effectively facilitated by the need for a large ratio a$_{\rm out}$/a$_{\rm in}$ in dynamically stable triples.  The variations in the observed compact binary BS systems \citep[e.g.][]{carney05,mathieu09} could be explained via stochasticity in the mass transfer process from the outer tertiary companion, as discussed in \citet{portegieszwart19} (i.,e., sometimes they accrete in steady-state via a circumbinary disk toward a mass ratio of unity, sometimes the mass ratio is driven away from unity and one accretes much more than the other, sometimes the two accreting MS stars merge, etc.).

Importantly, any WD remnant in the inner binary should (post-merger) be over-massive relative to what would be expected from single star evolution alone.  Hence, it could appear to have formed from a much younger progenitor star than the outer tertiary companion, which offers a potential smoking gun of a prior merger event (although we note that this is not a unique signature of this specific channel) \citep[e.g.][]{2019arXiv191005335T}. 



\subitem ii) \textit{Accretion onto a MS-WD inner binary (i.e., $m_{\rm 1}$ $>$ $m_{\rm 3}$ $>$ $m_{\rm 2}$)}

In this scenario, the outer tertiary would overfill its Roche lobe after only one of the inner binary components has evolved to become a WD.  Hence, any CBD would form around a tight WD-MS inner binary.  
As shown in Figure~\ref{fig:silvia3}, we find that of order $\sim$ 1-5\% of triple star systems should evolve to produce a tight WD-MS inner binary and a MS star outer tertiary companion that will both overflow its Roche lobe at some point over the course of its stellar evolution while also maintaining dynamical stability.  This is about a factor of two smaller than we find for the WD-WD inner binary case (see below).

But what comes next?  The answer depends on the initial relative masses between the components of the inner binary.  
If the MS star is initially less massive than the WD, which occurs in 20\% of our models, and the tendency is indeed to accrete toward a mass ratio of unity via a CBD, then the MS star should grow in mass until it reaches that of the WD.  This would then produce a blue straggler in a compact binary with a WD companion (and outer tertiary WD).  Thereafter, we naively expect most of the accreted mass to be lost from the system due to the difficulties associated with growing the WD in mass, but this would require detailed numerical simulations to verify.  Indeed, the assumption that WDs do not accrete any material is most likely closest to the reality, given the very fine-tuned mass transfer rates needed for the WD to grow steadily and avoid expelling the accreted mass \citep{bours13}.  If, on the other hand, the WD in the inner binary is initially less massive than its MS star companion, then the WD should have a higher accretion rate via a CBD than the MS companion and we expect the WD to expel most of the accreted mass.  If mass transfer onto the WD can somehow be sustained and it grows in mass, then either the WD explodes before catching up to the MS star in mass (i.e., before achieving $q$ $\sim$ 1), or the WD does catch up and the relative accretion rates onto the WD and MS star flip.  
%

Under favourable conditions, when the WD can grow in mass, it could eventually explode.  If the naive expectation that any subsequent accretion proceeds while approximately maintaining $q$ $\sim$ 1, then we would naively expect a hypervelocity MS star to be produced, with a mass very close to that of the WD at the time of explosion.  If the WD explodes when its mass roughly exceeds the Chandrasekhar mass limit, this predicts an excess of hypervelocity MS stars with masses of $\sim$ 1.4 M$_{\odot}$ in observed samples of hypervelocity stars.  

We can calculate an upper limit for the ejection velocity of such hypothetical MS HVSs as follows.  An upper limit follows from the assumption that the CBD effectively dissipates all of the orbital energy and angular momentum of the inner MS-WD binary, bringing the pair into contact at the time of explosion of the WD.   Assuming an approximate mass-radius relation for MS stars of M/M$_{\odot} =$ R/R$_{\odot}$ \citep{maeder09} and using Equation 27 in \citet{nauenberg72} to estimate the WD radius (adopting a mean molecular weight per electron of 2), we obtain an upper limit for the ejection velocity of the MS star of $\sim$ 618 km s$^{-1}$.

By construction, the primary of the inner binary is the most massive companion, and hence should produce the most massive WD by single star evolution alone.\footnote{This assumes the primary's evolution is not truncated early by interactions with the secondary.  For example, this could produce He WDs with masses $<$ 0.5 M$_{\odot}$.}  
Thus, this scenario predicts a high incidence of CO (or even ONe) WDs in the inner binary, or at least a tendency toward more massive WDs.  
After the MS star in the inner binary has had a chance to evolve to become a WD itself, this scenario predicts an over-massive CO WD relative to what might be expected from single star evolution alone (at least for this particular scenario for which we assume $m_{\rm 1}$ $>$ $m_{\rm 3} >$ $m_{\rm 2}$).  Such an over-massive CO WD could then be a smoking gun of a prior accretion or merger event, since it could appear to have formed from a much younger progenitor star than the outer tertiary companion.  This would be observable via the temperature of the WD and its associated age calculated from theoretical WD cooling curves \citep[e.g.][]{gosnell14,gosnell15}. 

Alternatively, the inner binary pair could merge, most likely producing a giant star.  In particular, this channel could form either an over-massive blue or yellow straggler \citep[e.g.][]{leiner16}, or even a sub-sub-giant branch star \citep[e.g.][]{geller17a,geller17b}, in a long-period binary with a WD or MS companion.  The approximate distribution of predicted orbital periods for this mechanism are shown in the left panel of Figure~\ref{fig:silvia5}.  

\subitem iii) \textit{Accretion onto a WD-WD inner binary (i.e., $m_{\rm 1}$ $>$ $m_{\rm 2}$ $>$ $m_{\rm 3}$)}

This is the scenario considered in Section~\ref{prog}, and quantified using binary evolution models in Section~\ref{binevmodels}.  As shown in Figure~\ref{fig:silvia2}, we find that of order $\sim$ 5-10\% of triple star systems should evolve to produce a compact WD-WD inner binary and a MS star outer tertiary companion that will both overflow its Roche lobe at some point over the course of its stellar evolution while also maintaining dynamical stability between the inner and outer orbits.  As quantified in Figure~\ref{fig:silvia2}, the inner WD-WD binary could be sufficiently compact to orbit inside any CBD that forms around it, once the tertiary has evolved to overfill its Roche lobe.  

What comes next is unclear theoretically.  Under favourable conditions, if the WDs are able to retain the accreted mass, then they could grow in mass while maintaining a mass ratio close to unity, similar to the scenario considered in \citet{portegieszwart19} for a MS-MS inner binary (see sub-item (i)).  If the tendency is indeed for the mass ratio to tend toward unity and the WDs are able to continue growing in mass until reaching the Chandrasekhar mass limit, then this could produce a hypervelocity WD \citep{hansen03,justham09,geier15,brown15}.  This is because the surviving WD will be ejected at the instantaneous orbital velocity at the time of explosion (see below).  We might naively expect the WDs to both be near the Chandrasekhar mass limit at the time of explosion, producing a hypervelocity WD with a mass of $\sim$ 1.4 M$_{\odot}$. Whether or not this scenario is correct depends on the many uncertain details of not only the accretion process, but also the detonation event that triggers the SN explosion \citep[e.g.][]{shen18}.  For most of the available parameter space, however, we might expect most of the mass accreted by the WDs to be expelled via novae or dwarf novae eruptions, preventing them from growing to near the Chandrasekhar mass limit.  We would naively expect the corresponding observational signatures of these events to be different than for the single WD case, motivating the triple scenario considered here when trying to interpret the observed features of any anomalous novae and/or dwarf novae.  Such anomalous features could appear in the form of increased X-ray production or in the accretion signature itself.

We can calculate an upper limit for the ejection velocity of such hypothetical WD HVSs as done above for MS-WD inner binaries, which yields 
an upper limit for the ejection velocity of the surviving WD of $\sim$ 5150 km s$^{-1}$.

By construction, the inner binary initially contains the most massive components of the system.  Hence, considering only single star evolution and ignoring any effects coming from the evolution of the inner binary (e.g., mass transfer and subsequent truncated stellar evolution, etc.), this scenario predicts the highest frequency of ONe WDs in the inner binary, or at least a tendency toward the inner binary hosting more massive CO WDs.  This scenario also predicts the smallest mass ratios q relative to the mass of the outer tertiary companion.  This is because binary evolution in the inner binary can truncate stellar evolution early, but this would not be the case for the outer tertiary companion which we assume reaches the AGB and hence has a core mass decided by single star evolution.

Alternatively, mass transfer from the outer tertiary companion could drive the inner binary pair to merge or, more likely, collide with non-zero orbital eccentricity \citep[e.g.][]{2020ApJ...891..160C}.   Hydrodynamical simulations suggest that little mass ($\lesssim$ 10$^{-3}$ M$_{\odot}$) should be lost during the merger process, such that the final WD mass should be roughly equal to the sum of the masses of the progenitor WDs \citep{aguilar09}.  Both WDs should have masses populating the more massive end of the WD mass distribution.  
Hence, such a merger would be more likely to produce a Type Ia SN event, relative to other scenarios (see below).  Alternatively, the merger product could collapse into an ONe WD or even a neutron star.  
As in the other scenarios, any CO/ONe WD remnant should (post-merger) be over-massive relative to what would be expected from single star evolution alone.  Hence, it could appear to have formed from a much younger progenitor star than the outer tertiary companion, which again offers a potential smoking gun of a prior accretion or merger event.



\item Our results can be used to calculate corresponding rates for the production of the three different triple scenarios described above.  This is done assuming a realistic initial population of triples, binaries and singles in the Galaxy, and computing a corresponding total stellar mass M$_{\rm tot}$ in singles, binaries and triples.  This M$_{\rm tot}$ can then be multiplied by a realistic star formation rate (SFR) to obtain a number of systems formed per year, which in turn should be multiplied by the fraction of systems in our simulated triple populations (relative to the assumed \textit{initial} total number) simultaneously satisfying both the Roche lobe-overflow and the dynamical stability constraints.  Note that these rates are computed using the first and second sets of numbers shown in the right panels of Figures~\ref{fig:silvia4}, ~\ref{fig:silvia3} and~\ref{fig:silvia2}, and therefore make the same assumptions as described in the figure captions.  
We adopt binary and triple fractions of, respectively, 40\% and 10\% \citep{raghavan10,toonen17}, along with a SFR of 3 M$_{\odot}$ yr$^{-1}$.  All calculations are performed assuming a tertiary mass of 1 M$_{\odot}$, such that our rate calculations correspond to lower limits (in terms of the requirement to be Roche-lobe filling at a given outer orbital separation), as described in preceding sections.  We note that including lower mass tertiaries would likely reduce our computed rates slightly, since drawing from a Kroupa initial mass function \citep{kroupa93} would reduce the rates by at most a factor of ten but observations suggest that the tertiary masses are much less likely to be this massive \citep{tokovinin18}.  This effect is counter-balanced by the fact that we do not adopt a prior assumption on the orbital separation distribution for the outer tertiary orbit, which further contributes to making our rate calculations lower limits.  This is because the observed distribution is approximately flat in 1/a$_{\rm out}$ \citep[e.g.][]{tokovinin18,toonen18}, which emphasizes shorter-period outer orbits and should contribute to an increase in our computed rates by about an order of magnitude.  

For triples hosting MS-MS, MS-WD and WD-WD inner binaries, we find lower limits to the production rates in units of number of systems per year of, respectively, $\sim$ 6.0 - 12 $\times$ 10$^{-4}$ ($\sim$ 42 - 85 $\times$ 10$^{-4}$) yr$^{-1}$, $\sim$ 1.5 - 3.1 $\times$ 10$^{-4}$  ($\sim$ 0.6 - 1.1 $\times$ 10$^{-4}$) yr$^{-1}$, and $\sim$ 0.7 - 1.4 $\times$ 10$^{-4}$ ($\sim$ 0.9 - 1.7 $\times$ 10$^{-4}$) yr$^{-1}$.  For the MS-MS binaries, the first rates consider only initial primary star masses in the range 0.95 - 10 M$_{\odot}$, whereas the second rates in parentheses assume initial primary masses in the range 0.1 - 100 M$_{\odot}$.  For the MS-WD and WD-WD cases, the first rates corresponds to our fiducial set of models and the rates in parentheses correspond to our enhanced CE prescription models.  Note that the rates are roughly independent of our chosen set of models (i.e., fiducial or enhanced CE prescription).  These estimates correspond to lower limits, since we assume a low-mass limit of 1 M$_{\odot}$ for the tertiaries, and more massive tertiaries would be Roche lobe overfilling at longer orbital periods.  Thus, on average, we expect a handful of candidate systems to appear in our Galaxy roughly every 10$^4$ years or so, with the rates being very similar for the MS-MS, MS-WD and WD-WD cases, but slightly higher for especially the MS-MS case but also the MS-WD case, relative to the WD-WD case.  For comparison, \citet{toonen12} simulate the full evolution of a population of triples combining stellar evolution and three-body dynamics.  These authors find a rate of tertiaries filling their Roche lobes before the inner MS-MS binary experiences mass transfer of 3 - 6 $\times$ 10$^{-4}$ yr$^{-1}$, which is in excellent agreement with our own estimate.  Similarly, \citet{nelemans01} find a total birth rate for WD-WD binaries of 3.2 $\times$ 10$^{-2}$ yr$^{-1}$ (Model C in their Table 2), such that the rate presented in this paper for WD-WD inner binaries in triples that can accommodate a CBD constitutes of order a percent of the total population of tight WD-WD binaries.

\item In all of the triple evolution scenarios considered here, any WDs must accrete at a stable finely tuned rate of $\sim$ 10$^{-7}$ M$_{\odot}$ yr$^{-1}$ in order for them to grow appreciably in mass \citep[e.g.][]{2007ApJ...663.1269N,2017gacv.workE..56K,2013ApJ...777..136W,bours13,shara17,shara18,2019MNRAS.490.1678C,2020NatAs.tmp...61H}.  This is because accretion onto a WD is expected to drive thermonuclear runaways on its surface for most mass transfer rates, producing novae and/or dwarf novae eruptions characteristic of cataclysmic variables (see \citet{bours13}, for example, for an overview).  Hence, many of the scenarios considered here involving the detonation of a WD could be relatively infrequent, depending on the poorly understood details of the mass transfer process.  However, if the WD does not explode, this could instead predict that novae and/or dwarf novae should be associated with the relevant triple scenarios considered in this paper (instead of Type Ia SN).  

\item Based on the above, we conclude that the contribution to the observed SN Ia rates from the mergers of WD-WD binaries via our mechanism should be relatively insignificant, and cannot alone account for the calculated disparity between the observed rates and those expected theoretically to occur from the double degenerate scenario \citep[e.g.][]{ruiter20}.  This is because we find that at most a few percent of our initial sample of MS triples could evolve to produce WD-WD inner binaries sufficiently compact to accrete in steady-state from a CBD due to mass transfer from an outer tertiary companion.  Moreover, as described above, the conditions required for WDs to experience stable accretion, and hence to grow appreciably in mass, may be difficult to satisfy.

\item As briefly touched upon above, the triple mass transfer scenarios considered here could contribute to the production of hypervelocity MS stars and WDs.  In particular, if the conditions required for steady-state accretion from a CBD are satisfied, we predict an excess of both WDs and MS stars with masses close to the Chandrasekhar mass limit in observed samples of Galactic HVSs \citep[e.g.][]{brown15,brown18,shen18}.  Given the results from our binary evolution models, we expect at most a few percent of an initial population of MS triples to evolve to produce systems capable of generating hypervelocity MS stars and WDs.  However, given that the expected masses for the MS HVSs are predicted here to be around $\sim$ 1.4 M$_{\odot}$, they should be observable out to at least some parts of the Galactic halo \citep[e.g.][]{brown18}.  The predicted hypervelocity WDs considered here likely would need to be within the Solar neighborhood to be detectable \citep{shen18}, making it difficult to achieve the sample size needed to actually observe in a sample of hypervelocity WDs any excess produced via the triple scenarios considered in this paper. 

\item When will the inner binary merge?  Naively, we might expect a higher merger rate for triples with retrograde configurations between their inner and outer orbits.  This is because any CBD that forms, upon interacting with the inner binary, would apply a torque that opposes the binary orbital motion.  This would most likely tend to remove angular momentum from the inner binary, either by increasing its eccentricity and/or decreasing its orbital separation, as observed in the simulations of \citet{portegieszwart19}.  This is because the torques should be maximized at apocentre, where the relative velocity between binary companions and the CBD is at a minimum.  This predicts over-massive BSs in long-period binaries with WD companions, with the rotation axis of the BS being oriented retrograde relative to the final binary orbit (originally the outer triple orbit).  
Naively, we expect prograde mergers to produce more massive BSs with larger rotation rates relative to retrograde mergers.  Although touched upon briefly in \citet{portegieszwart19}, this can and should be more robustly checked with numerical simulations.  This is because, if the CBD orbits prograde relative to the inner binary (which would occur if the tertiary orbit is prograde, by angular momentum conservation), any accretion from the CBD onto a prograde inner binary should come along with a higher accretion rate, given the lower relative velocities between the inner binary companions and the CBD.  Regardless, this makes an observationally testable prediction, specifically looking for a bimodality in the distribution of BS masses in long-period BS binaries.  Using observations of MS triples to determine what fraction of initial configurations are retrograde could therefore offer a method to constrain the physics of the mass transfer process, once weighed against the contribution of our proposed mechanism to observed long-period WD-BS systems \citep{mathieu09,gosnell14}.

Interestingly, some over-massive BSs have been observed, in particular in the old open cluster M67.  S1082 is a posited triple star system thought to contain two blue stragglers \citep{vandenberg01,sandquist03}, one in the inner compact binary ($P$ $\sim$ 1.07 days) with a mass of $\sim$ 2.7 M$_{\odot}$ and the other the outer tertiary ($P$ $\sim$ 1189 days).  S1237 is a long-period ($P$ $\sim$ 698 days) binary containing a yellow straggler, thought to be an evolved BS star \citep{leiner16} with a mass of 2.9 $\pm$ 0.2 M$_{\odot}$.  The companion in S1237 is thought to be a MS star near the turn-off, and possibly even a second BS.  The over-massive BSs in both S1082 and S1237 have masses over twice that of the MS turn-off in M67.  Given the nature of the companions in these systems, it seems unlikely that they were produced via the mechanism considered in this paper.  However, we note that a dynamical interaction could be needed to explain their origins, which could have acted to exchange out of the system a WD tertiary \citep[e.g.][]{hurley05}.  The timescales for such interactions (i.e., binary-binary, single-triple, binary-triple, etc.) to occur in M67 are of order $\sim$ 100 Myr \citep{leigh11,leiner16}, which is much smaller than the cluster age.  Hence, a scenario involving a previous dynamical exchange is entirely possible.  

Finally, we compare our predicted BS orbital periods to those derived by \citet{carney05} for a sample of field blue straggler binaries.  Their sample predicts mostly BS binaries with periods in the range of a few times $\sim$ 100-1000 days, as predicted by the triple scenario considered in this paper.  Thus, the mechanism proposed here naturally explains the clustering in orbital periods for observed BS binaries, under the assumption that the secondaries are WDs (the authors note that their observations are consistent with an unseen WD companion for all BS binaries).




\end{itemize}

\subsection{Triples with inner binaries containing NSs and/or BHs}


The scenario considered in this paper predicts mergers of near equal-mass WDs, leading to an interesting phenomenology 
including the progenitors of Type Ia supernovae,  along with the presence of an outer  tertiary companion.  However,
while here we specialized to low-mass stars for the more quantitative aspects of the analysis, triples can be formed with MS stars from any mass combination.  
If the inner binary is composed of at least one high-mass MS star, with $M$ $\gtrsim$ 8-10 M$_{\odot}$, then the compact object left behind
by that star would be a NS, or a BH for MS masses larger than $\approx$ 35-40 M$_{\odot}$  (e.g.  \citealt{Heger2003}). 
A setting with NSs/BHs as members of the inner binary can lead to a novel phenomenology and new astrophysical constraints,
as discussed in the following.

Let us start by considering the case where the inner binary is made up of two NSs.  The initial mass of each NS depends on the mass of the core of the progenitor star at the time of the collapse, as well as on the amount of fallback from the envelope.  However,  as the two NSs accrete from the CBD, we naively expect them to tend to achieve a similar mass over time. 
Depending on the orbital parameters of the inner binary, and the mass (and hence lifetime) of the outer tertiary star, one can envisage two scenarios. If the timescale for merger due to  gravitational  energy loss is shorter than the time that
it takes for the NS to reach a critical mass and collapse to a BH, then the merger will be one of two NSs of comparable mass.
This is expected to yield very little in terms of tidally disrupted ejecta (e.g. \citealt{Rezzolla2010}), and hence of prompt
electromagnetic (EM) counterpart shortly following  the GW signal 
(i.e. a "standard" short gamma-ray burst, such as the case of GRB170817A, detected 1.7~s after the gravitational wave signal
GW170817, \citealt{Abbott2017}). 

However, the presence of a disk surrounding the newly formed BH could potentially give rise to a novel phenomenology. 
If the (formerly) tertiary star remains bound, on a long timescale over which mass transfer continues,
the new system would be similar to that of an X-ray binary (XRB, which would be low or high-mass depending on the mass of the companion star), and hence be possibly detectable as an X-ray source.  X-ray luminosities for XRBs are observed up to $L_X\sim$~a few $\times 10^{39}$~erg~s$^{-1}$ \citep{Kim2004}. 

With the inner binary composed of a double NS accreting from a CBD, another interesting outcome can 
further be envisaged. 
A continuous period of accretion, with the NSs  already close to the maximum mass allowed by their equation
of state (EoS), could in fact be potentially interesting if one of the NS's accretes to surpass the critical value
for collapse to a BH before it merges with the other NS. This phenomenon  of accretion induced collapse, 
which has been studied via simulations in full GR by 
\citet{Giacomazzo2012}, has been found to be followed by a phase of rapid accretion onto the newly formed BH. 
The post-collapse accretion rates onto the BH, on the order of 10$^{-2}$ M$_{\odot}$~s$^{-1}$, would lead to a phenomenology
similar to that of $\gamma$-ray bursts, but happening in a triple system. If the system is not disrupted by the explosion,
it would now have an inner binary composed of an NS and a BH, with an outer evolved star. Since the NS of the inner binary   
would have a very comparable mass to the NS companion at the time of the accretion induced collapse, mass measurements
of the surviving NS could potentially be used to constrain the maximum NS mass. 

Additional constraints on the NS mass could come from consideration of WD-NS inner binaries accreting from a CBD.  If the WD should be less massive than the NS initially, it should accrete more from the CBD and grow faster in mass, provided it is able to retain the accreted material.  If the WD is able to reach the Chandrasekhar mass limit and detonate via a supernova event, this could produce a hypervelocity NS remnant.  
If the NS is spun up by the accretion process from the CBD, then these HVS NSs could be detectable as hypervelocity millisecond pulsars with masses near the Chandrasekhar mass limit. 

If on the other hand the outer binary star becomes unbound when the inner binary merges, and at least
a fraction of the CBD particles still remain bound, then
this configuration has the potential to lead to an electromagnetic counterpart following the GW signal from the binary merger.
This is an especially interesting situation, since it would similarly apply, at least at a qualitative level, to NS-NS, NS-BH, and
BH-BH mergers. For the NS-NS case, there would be a signal even for equal mass NSs, which are not expected to have any
significant post-merger emission from the tidally disrupted material at the merger, as already discussed above.   For an inner binary composed of an NS-BH, an EM signal is only expected for mass ratios smaller than $q$ $\sim$ 3-5 (where the exact value depends on the EoS of the NS), since for larger mass ratios the tidally disrupted NS would be swallowed by the BH without being able to form an accretion disk. Last, for a BH-BH merger, no electromagnetic signature is expected in the most common scenario of their merger
in the interstellar medium.  The presence of a circumbinary disk at the time of the merger could change any of the above
pictures in that it could lead to an EM signal when not expected. As such, it could be a telltale signature of the triple
evolutionary scenario presented in this paper.

The characteristics of the EM signature from a binary merging while surrounded by a circumbinary disk  
 were  discussed by \citet{deMink2017}. The binary orbital separation decays due to loss of gravitational
 wave energy on a timescale $t_{\rm GW}$. The evolution of the disk depends on the relative magnitude
 between $t_{\rm GW}$ and the viscous timescale at the inner disk $t_{\rm visc}$($R_{\rm in}$) 
 \citep{Milos2005,Perna2016,deMink2017}.  As long as 
 of the disk quickly readjusts to the decreasing orbital separation, and correspondingly fills the inner cavity.
 However, once the equation is reversed, the disk can no longer adapt to the rapid change, and its inner radius is 
 roughly determined by the value it had when $t_{\rm visc}$($R_{\rm in}$) $=$ $t_{\rm GW}$.  
 The particles of the CDB which still remain bound find themselves on elliptical orbits in a cool and thin disk. 
 Shocks are likely to ensue and yield potentially detectable EM signatures  
 (e.g. \citealt{Lippai2008, Schnittman2007,Schields2008,Corrales2010}). 
 Heating by the shocks
 occurs on a timescale on the order of the dynamical time \citep{deMink2017}, which hence yields the delay
$t_{\rm delay}$ $\sim$ $t_{\rm dyn}$  between the energy deposition in the disk and the GW merger signal
 \begin{equation}
t_{\rm  delay} \sim \frac{GM}{v^3} \sim 20 \frac{(M_{\rm bin}/10 M_{\odot})}{[v/(10^3 {\rm km/s})]^3}\;\; {\rm min},
 \end{equation}
where $M_{\rm bin}$ is the mass of the binary, and $v$ is the larger between the recoil velocity imparted to the
center of mass during the merger, and the Keplerian velocity at the inner orbit.
The luminosity of the event is the result of the dissipation of the kinetic energy over the timescale $t_{\rm dyn}$, and hence
is estimated as \citep{deMink2017}
\begin{equation}
L \sim \frac{M_{\rm disk}v^2}{t_{\rm dyn}} \sim 5 \times 10^{42} \left(\frac{f}{0.1}\right)\left( \frac{q_{\rm d}}{10^{-3}} \right) \left( \frac{v}{10^3 {\rm km/s}} \right)^5 \;\; {\rm erg}\; {\rm s}^{-1}\,,
\end{equation}
where $q_{\rm d}$ is the ratio between the disk mass and the binary mass and $f$ is a scaling factor calibrated
against simulations. The mass which is present in the CBD at the time of the merger, and which remains
bound, is a highly uncertain quantity, and hence the scalings are best given  in a parameterized
way. The characteristic temperature associated with this EM signal (the shock temperature) peaks 
around a few $\times$ 10$^7$~K, hence yielding a signal peaking in medium-energy X-rays.  

In summary, the considerations above show
 the potentially interesting phenomenology associated
with a triple system in which the inner binary is made up of any combination of WDs, NSs and BHs.  We defer to a future paper a more detailed analysis of outcomes involving NSs and BHs, given the larger parameter space and the significant number of uncertainties, including the added complication of, and uncertainties related to, natal kicks.  For now, we simply point out that we know of no other mechanism capable of producing mergers of CO binaries with mass ratios equal to unity.  Hence, if such objects are detected by gravitational wave detectors and confirmed to have very finely tuned mass ratios, it could be worth considering the triple evolution scenarios studied in this paper.  


\section{Summary} \label{sect:conclusions}

In this paper, we consider the formation of twin (i.e., with mass ratios near unity) compact binaries, with a focus on white dwarfs.  These systems may form through mass transfer from an outer Roche-lobe filling tertiary star. Once this star evolves off the main-sequence, part of its envelope is transferred to the inner binary, and accreted via a circumbinary disk by the two inner stars.  As illustrated in \citet{portegieszwart19}, the mass transfer stream could form a circumbinary disk, from which the inner binary stars accrete, potentially driving the pair toward a mass ratio close to unity.  The inner binary orbital separation can decrease or expand depending on the details of the transfer of mass and angular momentum.  Using analytic methods, we constrain the observed properties of the resulting WD-hosting triples and make predictions for the period of the outer orbit.  Our results for the predicted period distribution can further explain the observed features of the period distribution of BS-hosting binaries in the Galactic field \citep{carney05}.  


We quantify the expected frequency of triple star systems that could contain an inner MS-MS, MS-WD or WD-WD binary that can accommodate a circumbinary disk and via it accretion from the outer tertiary companion.  Specifically, we incorporate population synthesis-based calculations using binary evolution models performed with the \texttt{SeBa} code \citep{portegieszwart96,toonen12} to calculate, for a reasonable initial population of MS-MS binaries, the expected frequency of MS-MS, MS-WD and WD-WD compact binaries that could host outer triple companions that will evolve to become simultaneously Roche lobe over-filling while maintaining dynamical stability in both the inner and outer orbits.  
The expected frequency of systems that could satisfy both constraints is of order $\sim$ 1\%, $\sim$ 1-5\% and 5-10\% for, respectively, MS-MS, MS-WD and WD-WD binaries, nearly independent of our assumptions for the initial orbital properties of the MS-MS binaries.  We further present the predicted distributions of outer orbital periods for these scenarios.  Our results suggest that of order $\gtrsim$ 10\% of all primordial MS triples should evolve to be able to accommodate a CBD around the inner binary, given a realistic empirically-motivated population of initial triples.

Finally, using the expected frequencies computed above, we further calculate lower limits to the Galactic event rates for each of our three scenarios.  On average, we expect at least a handful of candidate systems to appear in our Galaxy roughly every 10$^4$ years or so, with the rates being very similar for the MS-MS, MS-WD and WD-WD cases, but slightly higher (by a factor of a few) for especially the MS-MS case but also the MS-WD case, relative to the WD-WD case.

These systems represent candidate blue stragglers, Type Ia SN and cataclysmic variables, via several different mass transfer channels and binary evolutionary pathways quantified here.  More broadly, the mechanism considered in this paper predicts mergers of twin binaries (i.e., with $q$ $\sim$ 1) containing MS stars and/or WDs.  We have argued that accretion onto these inner binaries via a CBD could produce exotic systems including blue stragglers, Type Ia SN, cataclysmic variables, etc. that typically come along with strong predictions for the observed properties of the system, including a narrow range of periods for the outer tertiary WD.  The mechanism proposed here could also produce hypervelocity MS stars, WDs, and even NSs and millisecond pulsars, with masses near the Chandrasekhar mass limit. We have discussed extending our results to the case of inner binaries containing NSs and BHs.  In this case, we expect the proposed triples to produce a signature that in at least some cases could be detectable using existing and/or future gravitational wave observatories, and that could be accompanied by unique EM counterparts relative to other known mechanisms.  This further motivates future hydrodynamics simulations, with a focus on the conditions needed to accumulate significant mass in the CBD, such that the proposed EM counterparts could actually be observable. 

\section*{Acknowledgments}

N.W.C.L. acknowledges support from Fondecyt Iniciac\'ion grant 11180005.  ST acknowledges support from the Netherlands Research Council NWO (VENI 639.041.645 grants).  SPZ would like to thank Norm Murray and CITA for their hospitality during his long-term visit.  This work was supported by the Netherlands Research School for Astronomy (NOVA). 
RP acknowledges support by NSF award AST-1616157.
In this work we use the matplotlib
\citep{2007CSE.....9...90H}, numpy
\citep{Oliphant2006ANumPy}, AMUSE
\citep{portegies_zwart_simon_2018_1443252}, SeBa
\citep{2012ascl.soft01003P}, Huayno \citep{2012NewA...17..711P}, MESA
\citep{2010ascl.soft10083P}, and GadGet2 \citep{2000ascl.soft03001S}
packages. The calculations ware performed using the LGM-II (NWO grant
\# 621.016.701) and the Dutch National Supercomputer at SURFSara
(grant \# 15520).

\bibliographystyle{mnras}
\bibliography{mnras}


\bsp

\label{lastpage}

\end{document}